\documentclass{aa}
\usepackage{graphics}
\def\dd{{\rm d}}
\def\zm{\zeta_{\rm m}}

\def\micron{\mu m}
\def\tE{t_{\oplus}}

\def\tco{t_{\rm co}}
\def\rg{r_{_\Gamma}}
\def\rG{r_{_\Gamma}}
\def\xiB{\xi_{_B}}
\def\xie{\xi_e}
\def\Eint{E_{\rm int}}
\def\Vco{V_{\rm co}}
\def\Macc{M_{\rm acc}}  % the swept-up mass
\def\ee{{\cal E}}
\def\eemin{{\cal E}_{\rm min}}

\def\eecool{{\cal E}_{\rm cool}}
\def\egcool{\gamma_{\rm cool}}

\def\nuEm{\nu_{\oplus, {\rm m}}}
\def\num{\nu_{\rm m}}

\def\nuabs{\nu_{\rm abs}}
\def\sigmaT{\sigma_{_T}}
\def\nucooE{\nu_{\oplus,{\rm cool}}}
\def\nucoo{\nu_{\rm cool}}
\def\nce1{\nu_{{\rm cool},\oplus, 1}}
\def\nce2{\nu_{{\rm cool},\oplus, 2}}
\def\gram{\, \hbox{g}}
\def\FEm{F_{\nu, \oplus, m}}
\def\FE{F_{\nu, \oplus}}
\def\brsub{b}
\def\tEi{t_{\oplus,\brsub}}
\def\Hz{\hbox{Hz}}
\def\cm{\hbox{cm}}
\def\erg{\hbox{erg}}
\def\day{\, \hbox{day}}
\def\Rhost{R_H}
\def\rad{\, \hbox{rad}}
\def\mag{\, \hbox{mag}}
\def\sterad{\, \hbox{Sr}}
\def\dl{d_{_{\rm L}}}
\def\Gpc{\, \hbox{Gpc}}

\begin{document}

\thesaurus{(13.07.1)}
\title{Gamma Ray Burst Beaming Constraints from Afterglow Light Curves}
% \title{A Beaming Limit from the GRB 970508 Afterglow}
% \subtitle{GRB 970508}
\author{James E. Rhoads}
\institute{Kitt Peak National Observatory}
\date{Received 24 December 1998 / Accepted }
\maketitle

\begin{abstract}
The beaming angle $\zm$ is the main uncertainty in gamma ray burst
energy requirements today.
We summarize predictions for the light curves of beamed bursts, and
model the R band light curve of GRB 970508 to derive 
$\zm \ga 30^{\degr}$.  This yields an irreducible minimum energy requirement
of $3.4 \times 10^{49}$ ergs to power the afterglow alone.
\end{abstract}

\section{Beamed Gamma Ray Burst Afterglow Models}
In the Rome meeting I presented a derivation of the dynamical behavior
of a beamed gamma ray burst (GRB) remnant and its consequences for the
afterglow light curve.  (Cf.\ Rhoads 1999 [Paper I]).  Here,
I summarize these results and apply them to test the range of beaming
angles permitted by the optical light curve of GRB 970508.

Suppose that ejecta from a GRB are emitted with initial Lorentz factor
$\Gamma_0$ into a cone of opening half-angle $\zm$ and expand into an
ambient medium of uniform mass density $\rho$ with negligible
radiative energy losses.  Let the initial kinetic energy and rest mass
of the ejecta be $E_0$ and $M_0$, and the swept-up mass and internal
energy of the expanding blast wave be $\Macc$ and $\Eint$.  Then
energy conservation implies $\Gamma \Eint \approx \Gamma^2 \Macc c^2
\approx E_0 \approx \hbox{constant}$ so long as $1/\Gamma_0 \la \Macc
/ M_0 \la \Gamma_0$.

The swept-up mass is determined by the working surface area: $\dd
\Macc / \dd r \approx \pi (\zm r + c_s \tco)^2$, where $c_s$ and
$\tco$ are the sound speed and time since the burst in the frame of
the blast wave $+$ accreted material.  Once $\Gamma \la 1/\zm$, $c_s
\tco \ga \zm r$ and the dynamical evolution with radius $r$ changes
from $\Gamma \propto r^{-3/2}$ to $\Gamma \propto \exp(-r/\rg)$
(Rhoads 1998, \& Paper I).  The relation between observer frame time
$\tE$ and radius $r$ also changes, from $\tE \propto r^{1/4}$ to $\tE
\propto \exp(r/[2 \rG])$.  Thus, at early times $\Gamma \propto
\tE^{-3/8}$, while at late times $\Gamma \propto \tE^{-1/2}$.  The
characteristic length scale is $\rg = \left(E_0 / \pi c_s^2 \rho
\right)^{1/3}$, and the characteristic observed transition time
between the two regimes is $\tEi \approx 1.125 (1+z) \left( E_0 c^3 /
[\rho c_s^8 \zm^2] \right)^{1/3} \zm^{8/3}$, where $z$ is the burst's
redshift.

We assume that swept-up electrons are injected with a power law energy
distribution $N(\ee) \propto \ee^{-p}$ for $\ee = \gamma_e m_e c^2 >
\eemin \approx \xie m_p c^2 \Gamma$, with $p > 2$, and contain a
fraction $\xie$ of $\Eint$.  This power law extends up to the cooling
break, $\eecool$, at which energy the cooling time is
comparable to the dynamical expansion time of the remnant.  Above
$\eecool$, the balance between electron injection (with
$N_{\hbox{inj}} \propto \ee^{-p}$) and cooling gives $N(\ee) \propto
\ee^{-(p+1)}$.

We also assume a tangled magnetic field containing a fraction $\xiB$
of $\Eint$.  The comoving volume $\Vco$ and burster-frame volume $V$
are related by $\Vco \approx V/\Gamma \propto \Macc / \Gamma$, so that
$B^2 = {8 \pi \xiB \Eint / \Vco} \propto \Gamma^2$ and $B \propto \Gamma$.

The resulting spectrum has peak flux density $\FEm \propto \Gamma B
\Macc/ \max(\zm^2, \Gamma^{-2})$  at an observed frequency $\nuEm \propto \Gamma B
\eemin^2 / (1+z) \propto \Gamma^4 / (1+z)$.  Additional spectral
features occur at the frequencies of optically thick synchrotron self
absorption (which we shall neglect) and the cooling frequency
$\nucooE$ (which is important for optical observations of GRB 970508).
The cooling break frequency follows from the relations $\egcool \approx (6 \pi
m_e c)/ ( \sigmaT \Gamma B^2 \tE)$ (Sari, Piran, \& Narayan 1998;
Wijers \& Galama 1998) and $\nucooE \propto \Gamma B \egcool^2 \propto
(\Gamma^4 \tE^2)^{-1}$.  In the power law regime, $\FEm \propto
\tE^{0}$, $\nuEm \propto \tE^{-3/2}$, and $\nucooE \propto
\tE^{-1/2}$; while in the exponential regime, $\FEm \propto \tE^{-1}$,
$\nuEm \propto \tE^{-2}$, and $\nucooE \propto \tE^0$.  The spectrum
is approximated by a broken power law, $F_\nu \propto \nu^{-\beta}$,
with $\beta \approx -1/3$ for $\nu<\nuEm$, $\beta \approx (p-1)/2$ for
$\nuEm < \nu < \nucooE$, and $\beta \approx p/2$ for $\nu > \nucooE$.

The afterglow light curve follows from the spectral shape and the time
behavior of the break frequencies. Asymptotic slopes are given in table~1.
For the $\Gamma \sim 1/\zm$ regime, we study the evolution of break 
frequencies numerically.  The results for $\nuEm$ and $\FEm$ are given 
in Paper~I.  For $\nucooE$, a good approximation is
\begin{eqnarray*}
\lefteqn{
\nucooE = \left[ 5.89 \times 10^{13} \left( \tE / \tEi \right)^{-1/2}
 + 1.34 \times 10^{14} \right] \Hz } \\
& \times &
\left(1 \over 1+z \right)
\left(c_s \over c/\sqrt{3} \right)^{17/6}
\left( \xiB \over 0.1\right)^{-3/2} \\
 & \times & \left( \rho \cdot \cm^{3} \over 10^{-24} \gram   \right)^{-5/6}
\left( E_0 / 10^{53} \erg \over \zm^2 / 4 \right)^{-2/3}
\left( \zm \over 0.1 \right)^{-4/3} ~~.
\end{eqnarray*}

% Note, time slope \alpha = 1/2 - 3 p / 4  above cooling break in early
%  time regime.  time slope = 3/4 - 3 p / 4 below cooling break in
%  early time regime.
\begin{table}
\begin{tabular}{llll}
 & $\nuabs < \nu < \num$ & $\num < \nu < \nucoo$ & $\nucoo < \nu$ \\
$\tE \ll \tEi$ & $1/2$ & $3/4 - 3 p / 4$ & $1/2 - 3 p / 4$ \\ 
$\tE \gg \tEi$ & $-1/3$ & $-p$ & $-p$ \\
\end{tabular}
\caption{Light curve exponents $\alpha$ as a function of frequency and
time.  Here $\FE \propto \tE^\alpha$.}
\end{table}

\section{Application to GRB 970508}
In the best-sampled GRB afterglow light curve yet available (the GRB
970508 R band data), the optical spectrum changed slope
at $\tE \sim 1.4 \day$, suggesting the passage of the cooling break
through the optical band (Galama et al 1998).  We explore
the range of acceptable beaming angles for this burst by
fitting the afterglow light curve for $1.3 \day \le \tE \le 95 \day$
assuming that $\nucooE < c/0.7 \micron$.

The range of acceptable energy distribution slopes $p$ for swept-up
electrons is taken from the optical colors.  Precise measurements for
$2 \day \la \tE \la 5 \day$ give $F_\nu \propto \nu^{-\beta}$ with
$\beta = 1.10 \pm 0.08$ (Zharikov, Sokolov, \& Baryshev 1998), so that
$p = 2.20 \pm 0.16$.  We take this value to hold throughout the
range $1.3 \day \le \tE \le 95 \day$, thus assuming that $p$ does not
change as the afterglow evolves.  We subtract the host galaxy flux
($\Rhost = 25.55 \pm 0.19$; Zharikov et al 1998) from all data points
before fitting.

% The light curve break at $\tEi$ is very gradual, and the agreement or
% disagreement between $p$ and the observed light curve decay is
% therefore the primary test of beaming.

\begin{table}
\begin{tabular}{llllll}
Model & $\Rhost$ & $p$ & $\lg(\tEi/\day)$ & $R_c(t_0)$ &
 $\chi^2/\hbox{d.o.f.}$ \\
1 & 25.74 & 2.36 & 9 & 20.32 & 4.34 \\ %177.9 for 41 d.o.f.
2 & 25.55 & 2.20 & 5 & 20.33 & 3.80 \\ %155.9
3 & 25.36 & 2.04 & 3.75 & 20.32 & 3.55 \\ % 145.7
\end{tabular}
\caption{Fitted break times $\tEi$ and magnitudes $R_c(t_0)$ (at
fiducial observed time $t_0 = 3.231$ day) for beamed GRB afterglow
models for three pairs of acceptable host galaxy magnitude $\Rhost$
and electron power law index $p$. The fit included all 43 data points
with $1.3 \day \le \tE \le 95 \day$ in the compilation by Garcia et al
(1998).}
\end{table}

\begin{figure}
  \resizebox{\hsize}{!}{\includegraphics{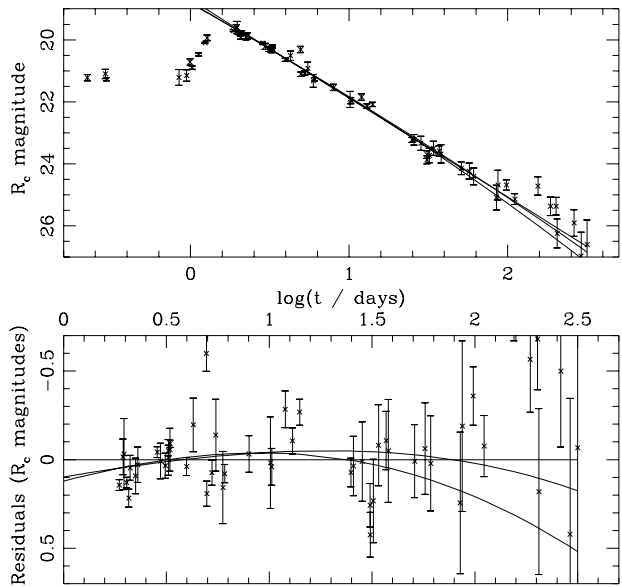}}
  \caption{Upper panel: The Cousins R band light curve for GRB 970508
  with the three fits shown in table 1.  Lower panel: Residuals for
  the data and for models 2 and 3 (in order of increasing curvature) 
  relative to model~1.  A host galaxy
  flux corresponding to $\Rhost=25.55$ has been subtracted from all
  data points. } \label{fig1}
\end{figure}

We fixed values of $\Rhost$ and $p$, and then executed
a grid search on the break time $\tEi$ and normalization of the model
light curve.  Results are summarized in table~2 and figure~1.  The
final $\chi^2$ per degree of freedom is $\sim 4$.

These large $\chi^2$ values make meaningful error estimates on
parameters difficult.  Let us suppose $\chi^2$ is large because
details omitted from the models (%e.g.,
clumps in the ambient medium
or blast wave instabilities) affect the light curve, and so attach an
uncertainty of $0.1 \mag$ to each predicted flux.
Adding this in quadrature to observational uncertainties when computing
$\chi^2$, we obtain $\chi^2 / \hbox{d.o.f.} \sim 1$.  Error estimates
based on changes in $\chi^2$ then rule out $\lg(\tEi/\day) < 3.5$ at
about the 90\% confidence level even for our ``maximum beaming'' case
($p=2.04$, $\Rhost = 25.36$).
%% Really, what we have is increase in chi^2 from 48.52 to 51.07 for
%% best models with log(\tEi) = 3.75 and 3.5 .
%% For  log(\tEi) = 4.00 we get chi^2 = 52.67.

% Conversion into beaming angle:
To convert a supposed break time $\tEi$ into a beaming angle $\zm$,
we need estimates of the burst energy per steradian and the
ambient density.   Wijers \& Galama (1998)  infer $E_0 / \Omega = 
3.7 \times 10^{52} \erg / (4 \pi \sterad)$ and
$\rho = 5.8\times 10^{-26} \gram/\cm^3$. %from the times and frequencies of
% the various spectral breaks in the GRB 970508 afterglow.  
Combining these values with $\tEi \ga 10^{3.5} \day$ gives
% $\zm \ga 0.51 \rad = 29\deg$.
$\zm \ga 0.5 \rad \approx 30\deg$.  $E_0 / \Omega$ and $\rho$
are substantially uncertain, but because $\zm \propto (\rho / E_0
)^{1/8}$, the error budget for $\zm$ is dominated by
uncertainties in $p$ rather than in $E_0$ or $\rho$.

This beaming limit implies $\Omega \ge 0.75 \sterad$, which is $6\%$
of the sky.  GRB 970508 was at $z \ge 0.835$ (Metzger et al 1997).  We
then find gamma ray energy $E_\gamma = 2.8 \times 10^{50} \erg \times
(\Omega / 0.75 \sterad) (\dl / 4.82 \Gpc)^2 \left( 1.835 /
[1+z] \right)$.  If the afterglow is primarily powered by different
ejecta from the initial GRB, as when a ``slow'' wind ($\Gamma_0 \sim
10$) dominates the ejecta energy, then our beaming limit applies only
to the afterglow emission.  The optical fluence implies $E_{\rm opt} =
3.4 \times 10^{49} \erg \times (\Omega / 0.75 \sterad) (\dl /
4.82 \Gpc)^2 \left( 1.835 / [1+z] \right)$.  The irreducible minimum
energy is thus $3.4 \times 10^{49} \erg$, using the smallest possible
redshift and beaming angle.  We have reduced the beaming uncertainty,
from the factor $\sim \Gamma_0^2 \sim 300^2 \sim 10^5$ allowed by
$\gamma$-ray observations alone to a factor $(4 \pi \sterad)/(0.75
\sterad) \sim 20$, and thus obtain the most rigorous lower limit on
GRB energy requirements yet.

\begin{acknowledgements}
I thank Re'em Sari for useful comments.
\end{acknowledgements}

\end{document}